\documentstyle[12pt,epsf]{article}
\def\beq{\begin{equation}}
\def\eeq{\end{equation}}
\def\sw{\sin^2 \theta_W}
\def\cw{\cos^2 \theta_W}
\def\ap#1#2#3 {Ann. Phys. (NY) {\bf#1} (19#2) #3}
\def\apj#1#2#3 {Astrophys. J. {\bf#1} (19#2) #3}
\def\apjl#1#2#3 {Astrophys. J. Lett. {\bf#1} (19#2) #3}
\def\app#1#2#3 {Acta. Phys. Pol. {\bf#1} (19#2) #3}
\def\ar#1#2#3 {Ann. Rev. Nucl. Part. Sci. {\bf#1} (19#2) #3}
\def\cpc#1#2#3 {Computer Phys. Comm. {\bf#1} (19#2) #3}
\def\err#1#2#3 {{\it Erratum} {\bf#1} (19#2) #3}
\def\ib#1#2#3 {{\it ibid.} {\bf#1} (19#2) #3}
\def\jmp#1#2#3 {J. Math. Phys. {\bf#1} (19#2) #3}
\def\ijmp#1#2#3 {Int. J. Mod. Phys. {\bf#1} (19#2) #3}
\def\jetp#1#2#3 {JETP Lett. {\bf#1} (19#2) #3}
\def\jpg#1#2#3 {J. Phys. G. {\bf#1} (19#2) #3}
\def\mpl#1#2#3 {Mod. Phys. Lett. {\bf#1} (19#2) #3}
\def\nat#1#2#3 {Nature (London) {\bf#1} (19#2) #3}
\def\nc#1#2#3 {Nuovo Cim. {\bf#1} (19#2) #3}
\def\nim#1#2#3 {Nucl. Instr. Meth. {\bf#1} (19#2) #3}
\def\np#1#2#3 {Nucl. Phys. {\bf#1} (19#2) #3}
\def\pcps#1#2#3 {Proc. Cam. Phil. Soc. {\bf#1} (#2) #3}
\def\pl#1#2#3 {Phys. Lett. {\bf#1} (19#2) #3}
\def\prep#1#2#3 {Phys. Rep. {\bf#1} (19#2) #3}
\def\prev#1#2#3 {Phys. Rev. {\bf#1} (19#2) #3}
\def\prl#1#2#3 {Phys. Rev. Lett. {\bf#1} (19#2) #3}
\def\prs#1#2#3 {Proc. Roy. Soc. {\bf#1} (19#2) #3}
\def\ptp#1#2#3 {Prog. Th. Phys. {\bf#1} (19#2) #3}
\def\ps#1#2#3 {Physica Scripta {\bf#1} (19#2) #3}
\def\rmp#1#2#3 {Rev. Mod. Phys. {\bf#1} (19#2) #3}
\def\rpp#1#2#3 {Rep. Prog. Phys. {\bf#1} (19#2) #3}
\def\sjnp#1#2#3 {Sov. J. Nucl. Phys. {\bf#1} (19#2) #3}
\def\spj#1#2#3 {Sov. Phys. JEPT {\bf#1} (19#2) #3}
\def\spu#1#2#3 {Sov. Phys. Usp. {\bf#1} (19#2) #3}
\def\zp#1#2#3 {Zeit. Phys. {\bf#1} (19#2) #3}

\begin{document}
\begin{titlepage}
\begin{center}
{\Large \bf Theoretical Physics Institute \\
University of Minnesota \\}  \end{center}
\vspace{0.2in}
\begin{flushright}
TPI-MINN-98/02-T \\
UMN-TH-1623-98 \\
hep-ph/9802234\\
\end{flushright}
\vspace{0.3in}
\begin{center}
{\Large \bf  Top quark longitudinal polarization near the threshold in
$\ell^+ \, \ell^-$ annihilation\\}
\vspace{0.2in}
{\bf B.M. Chibisov \\}
School of Physics and Astronomy, University of Minnesota, Minneapolis,
MN
55455 \\

\vspace{0.2cm}

and

\vspace{0.2cm}

{\bf M.B. Voloshin  \\ }
Theoretical Physics Institute, University of Minnesota, Minneapolis,
MN
55455 \\ and \\
Institute of Theoretical and Experimental Physics, Moscow, 117259
\\[0.2in]
\end{center}

\begin{abstract}

We show that the longitudinal polarization of the top quarks produced in
the annihilation of $e^+ e^-$ or $\mu^+ \mu^-$ into  $\bar t t$ at
energies near the threshold is not affected by the large Coulomb-type
corrections, which greatly modify the total cross section.
Thus the longitudinal polarization, although small, may provide an
independent information on the mass and the width of the top quark,
largely independent of the uncertainty in $\alpha_s$.

\end{abstract}
\end{titlepage}

It has been pointed out well some time ago$^{\cite{fk}}$ that due to the
large mass and width of the top quark, the process of annihilation of a
lepton pair ($\ell^+ \ell^-$) into $\bar t t$ is dominantly determined
by the perturbative QCD, while the nonperturbative effects are
effectively blocked by the large mass and the fast decay of the top,
even in the threshold region. Recently there has been a revival of
interest to this process$^{\cite{hoang,berger,jezabek}}$ related to the
prospects of the future Next Linear Collider (NLC) and/or First Muon
Collider (FMC). The region near the threshold of the $\bar t t$
production presents a special interest, since it provides the best
sensitivity of the data to the kinematical parameters (mass, width) of
the top quark as well as to the value of the strong coupling $\alpha_s$.
The reason for the sensitivity to $\alpha_s$ is the well-known behavior
of the perturbation theory series near the threshold, where the
Coulomb-like interaction between quarks, each having in the c.m.
nonrelativistic velocity $v$ is described by the parameter
($\alpha_s/v$). The other side of this behavior is that at $v \le
O(\alpha_s)$ one has to sum all the Coulomb terms of the type
$(\alpha_s/v)^n$ exactly. This summation is performed by using the exact
solution of the Schr\"odinger equation for the Green's function in the
Coulomb potential, and also the effects of the running of $\alpha_s$ can
be accounted for in the leading-log approximation as well as the first
radiative correction, thus effectively summing also the terms of the
form $\alpha_s \, (\alpha_s/v)^n$ (see e.g. Ref.\cite{mv}). As to the
next term in this expansion that includes the radiative effects in order
$\alpha_s^2$ as well as the relativistic corrections $O(v^2)$, these are
calculated most recently$^{\cite{ht}}$, following earlier results for an
Abelian theory$^{\cite{hoang}}$ and the results of a full QCD
calculation$^{\cite{chetyrkin,cm}}$ of the cross section near the
threshold in order $\alpha_s^2$, but without the summation of the
additional Coulomb factors with $(\alpha_s/v)^n$.

The sensitivity of the production cross section to both the parameters
of the top quark and the coupling constant $\alpha_s$ implies that there
might be a considerable correlation between the values extracted from
the data on the total cross section only. Thus it would be helpful to
incorporate additional data with a different dependence on these
parameters. It is the purpose of this letter to point out that the
longitudinal polarization of the top quark, although small in the near
threshold region, is not affected by the Coulomb factors or the running
of the coupling in the leading-log order. Thus it is dominantly
sensitive to the mass and the width of the top quark, and, if measured,
can be used as an independent input in the determination of the
parameters from the data. Additionally, the longitudinal polarization,
determined by the interference of the top quark axial and vector
couplings, may provide limits on possible deviations of these couplings
from the Standard Model values{\footnote{Certainly, the threshold region
does not offer much of advantage for measuring the top quark axial and
vector couplings, since the longitudinal polarization is more prominent
in the high-energy region$^{\cite{as}}$, provided that the latter region
would be accessible on future colliders}.

Our consideration  here is limited to the leading Coulomb terms of the
form $(\alpha_s/v)^n$, the effects of the running of the coupling in the
leading approximation and to the lowest order of the nonrelativistic
expansion. We find that in this order the longitudinal polarization does
not receive any QCD corrections. The only effect of order $\alpha_s$
comes from the known hard gluon correction to the vector and the axial
vertices at the threshold and is determined by $\alpha_s(m_t)$. The
unaccounted subsequent corrections are thus of the relative magnitude
$\alpha_s^2$, $\alpha_s \, v$ and $v^2$, with possible logarithms of
$v$. Also, strictly speaking, there are corrections due to electroweak
radiative effects, i.e. of order $\alpha$, which apriori are less than
the unaccounted QCD effects.

It should be noted that the longitudinal polarization of the top
quark, $P_L$, discussed here is the one averaged over the production
angle: $\langle P_L \rangle$, i.e. the one determined by the parity
violation on the top quark side, rather than on the lepton side (for
further details see e.g. Ref. \cite{as}).

The top quark polarization in the near threshold region was previously
studied$^{\cite{hjkt}}$ numerically within a potential model approach.
We believe that additional assumptions inherent in such approach are not
necessary, and the type of polarization discussed here can be calculated
analytically with acceptable accuracy within the perturbative QCD, while
the non-perturbative effects, presumably small, deserve a special study,
which is beyond the present paper. It can be also noted, that a somewhat
similar insensitivity of the longitudinal polarization to the
Coulom-type effects was observed in the calculation$^{\cite{fkk}}$ of
the top production in two photon collisions.

Proceeding to the details of our argument, we write the well known
expressions (see e.g. Ref. \cite{as}) for the total cross
section of the process $\ell^+\ell^- \to \bar t t$ and for the
longitudinal polarization of the $t$ in the form
\beq
\sigma_{tot}= { 2 \, \pi \, \alpha^2 \, v \over  s} \, \left \{ (3-v^2)
\, D_V \,
|V|^2 + 2 \, v^2 \, D_A \, |A|^2 \right \}~~,
\label{sigtot}
\eeq
\beq
\langle P_L \rangle =  - { 4 \, {\rm Re}\left[v \,  V \, A^* \, D_{VA}
\right]
\over (3-v^2) \, D_V \, |V|^2 + 2 \, v^2 \, D_A \, |A|^2 }~~.
\label{pl}
\eeq
The factors $D$ in these equations read in the Standard Model as
\beq
D_V=\left | (1-4\, \sw)\left(1-{8 \over 3}\, \sw \right) \, z + {2 \over
3} \right
|^2  +  \left | \left (1-{8 \over 3}\, \sw \right) \, z \right |^2~~,
\label{dv}
\eeq
\beq
D_A=\left [ (1-4 \, \sw)^2 +1 \right ] \, |z|^2~~,
\label{da}
\eeq
and
\beq
D_{VA}= (1-4 \, \sw) \left [ (1-4\, \sw)\left(1-{8 \over 3}\, \sw
\right) \, |z|^2
+ {2 \over 3}\, z^* \right ]  +\left(1-{8 \over 3}\, \sw \right) \,
|z|^2 ~~,
\label{dva}
\eeq
where $z$ is the ratio of the $Z$ and the photon propagators and also
includes the ratio of their couplings. If the electroweak radiative
effects in the propagators are ignored, this ratio has the form
\beq
z= {1 \over 16 \, \sw \, \cw}\,{s \over {s-m_Z^2}}~~.
\label{zdef}
\eeq

The factors $V$ and $A$ in the equations (\ref{sigtot}) and (\ref{pl})
stand for the renormalization of respectively the vector and the axial
top quark vertices by the QCD effects\footnote{These factors also
receive electroweak corrections both in their absolute values and the
complex phases, if these corrections were not neglected, they would have
to be included along with the corrections to the ratio $z$ in
eq.(\ref{zdef}).} and constitute the subject of primary concern in this
paper. In the nonrelativistic limit of small $v$ the equation (\ref{pl})
simplifies as
\beq
\langle P_L \rangle = - { 4\, D_{VA} \over 3 \, D_V} \, {\rm Re}\left (
v \,  {A \over V} \right )
\approx -0.20 \, {\rm Re} \left ( v \, {A \over V} \right )~~,
\label{spl}
\eeq
where the numerical value $\sw=0.232$ is used as well as $m_t=175 \,
GeV$ (i.e. the threshold is at $\sqrt{s}=350 \, GeV$).

The QCD contribution to the factors $V$ and $A$ in the threshold region
comes from two well separated sources: the hard gluon correction to the
appropriate vertices, determined by the momenta of the virtual gluon of
order $m_t$, and the Coulomb corrections, determined by an effective
virtual momenta scale, which is a combination of the actual quark
momentum $p=m_t \, v$ and the Bohr momentum $p_B=2\, m_t \,
\alpha_s/3$. The former correction is included in the expression for the
vector and the axial currents in terms of equivalent operators in a
nonrelativistic description of the $\bar t \, t$ pair at the threshold:
\beq
\left ( \bar t \, {\mbox{\boldmath $\gamma$}} \, t \right ) = \left ( 1-
{8\,
\alpha_s\over 3 \, \pi}\right ) \, \left( \tilde \chi \,
{\mbox{\boldmath
$\sigma$}} \,\phi \right) \delta({\mbox{\boldmath $r$}})~,~~~~
\left ( \bar t \, {\mbox{\boldmath $\gamma$}}\gamma_5 \, t \right )= -
\left ( 1- {4\, \alpha_s\over 3 \, \pi}\right ) \, \left( \tilde \chi \,
{\mbox{\boldmath $\sigma$}} \,\phi \right) \times {{\mbox{\boldmath
$\nabla$}}
\over m_t} \delta({\mbox{\boldmath $r$}})~,
\label{vertex}
\eeq
where $\phi$ and $\chi$ are the nonrelativistic spinors for the quark
and the antiquark ($\tilde \chi=\chi^T \, (i \, \sigma_2)$), and
$\mbox{\boldmath $r$}$ is the vector of their relative position. The
hard renormalization factor in the vector current is well known, while
the same in the axial current can be read off the results of Refs.
\cite{rry} and \cite{jlz}. Thus with the inclusion of the Coulomb
effects the factors $V$ and $A$ can be written as:
\beq
V=\left ( 1- {8\, \alpha_s\over 3 \, \pi}\right ) \, S ~,~~~~~
A=\left ( 1- {4\, \alpha_s\over 3 \, \pi}\right ) \, P~,
\label{spdef}
\eeq
where $S$ and $P$ describe the Coulomb renormalization of the wave
function near the origin for respectively the $S$ wave and the $P$ wave:
\beq
S={\psi_C(0) \over \psi_0(0)}~,~~~~
P= \left. {\mbox{\boldmath $\nabla$}\psi_C(\mbox{\boldmath $r$}) \over
\mbox{\boldmath $\nabla$}\psi_0(\mbox{\boldmath $r$})} \right
|_{\mbox{\boldmath $r$}=0}~,
\label{spform}
\eeq
with $\psi_0(\mbox{\boldmath $r$})$ and $\psi_C(\mbox{\boldmath $r$})$
being the free wave function and  the one in the Coulomb potential.

The appropriate wave functions for considering the processes where the
top quark has definite momentum $\mbox{\boldmath $p$}$ are the plane
wave for the free wave function: $\psi_0(\mbox{\boldmath $r$})=\exp (i\,
\mbox{\boldmath $p \cdot r$})$ and the scattering state wave function
$\psi^{(+)}(\mbox{\boldmath $r$})$ in the Coulomb potential, which
function at infinity is a superposition of the plain wave with the
momentum $\mbox{\boldmath $p$}$ and a scattered spherical
wave$^{\cite{ll}}$:
\beq
\psi_C(\mbox{\boldmath $r$})=e^{\pi  \lambda/2} \, \Gamma(1-i \,
\lambda)\, e^{i \, \mbox{\boldmath $p \cdot r$}} \, _1 F_1 (i \,
\lambda, \, 1, i \, p r - i \, \mbox{\boldmath $p \cdot r$})~,
\label{psic}
\eeq
where $\lambda=2 \alpha_s m_t/(3 p)=2 \alpha_s/(3 v)$. With these wave
functions one readily finds the factors $S$ and $P$ from the equations
(\ref{spform}):
\beq
S=e^{\pi  \lambda/2} \, \Gamma(1-i \, \lambda)~, ~~~~P=e^{\pi
\lambda/2} \, \Gamma(1-i \, \lambda)\, (1-i \, \lambda)~.
\label{spres}
\eeq
Using eq.(\ref{spdef}), one finds from eq.(\ref{spl}) a remarkably
simple final result for the longitudinal polarization near the
threshold:
\beq
\langle P_L \rangle=
-C \, \left ( 1+ {4\, \alpha_s\over 3 \, \pi}\right ) {\rm Re} \left ( v
- i \, {2 \over 3} \, \alpha_s \right )= -C \, \left ( 1+ {4\,
\alpha_s\over 3 \, \pi}\right ) \, v
\label{plres}
\eeq
with $C=4 D_{VA}/(3 D_V) \approx 0.20$.

It is important, that the expressions in eq.(\ref{spres}) give the
factors $S$ and $P$ with their phases. The correct phase relation
between the $S$ wave and the $P$ wave is fixed by using the scattering
state Coulomb wave function$^{\cite{ll}}$. In terms of the real (at real
$\mbox{\boldmath $p$}$) radial wave functions $R_{pl}(r)$ and the
scattering phases $\delta_l$ for the partial waves the ratio $P/S$ can
be written as
$$
{P \over S}= { R^{(C)\,\prime}_{p1} (0)/R^{(0)\,\prime}_{p1} (0) \over
R^{(C)}_{p0}(0)/R^{(0)}_{p0}(0)} \, e^{i (\delta_1-\delta_0)}~~.
$$
In terms of this formula the Coulomb effect on the ratio of the absolute
values $|P/S|$ is cancelled by that on the phase difference
$\delta_1-\delta_0$ in the real part Re$(P/S)$.

The absence of dependence of the discussed here $P-S$ interference term
on the Coulomb coupling implies that there is also no effect on this
term from the running of the coupling in the leading order. Indeed,
taking into account the running in the leading order for a quantity
$F(\alpha_s)$ reduces to replacing the coupling $\alpha_s$ by the
effective coupling at a scale $\mu$ determined by the process:
$F(\alpha_s) \to F(\alpha_s(\mu))$. In the case considered the effect of
the Coulomb interaction on the longitudinal polarization is absent, thus
the effect of the running of the coupling at the Coulomb scale does not
enter the result. This potentially leaves the discussed polarization
sensitive only to subsequent corrections, which are of the relative
order $\alpha_s^2$.

Another point to be considered for the realistic top quarks is related
to the top decay width $\Gamma$. Formally the decay width can be
included by shifting the energy above the threshold, $E = \sqrt{s} -
2m_t$, in the complex plain: $E \to E+i\Gamma$ (this approximation
neglect effects of the relative order $\Gamma/m_t$), resulting in a
complex shift of the velocity: $v=\sqrt{(E+i\Gamma)/m_t}$. The latter
shift can be readily accounted for in our consideration of the Coulomb
effects and in the final result in eq.(\ref{plres}) by keeping $v$ as a
complex parameter. Clearly, for a complex $v$ the equation (\ref{plres})
assumes the form
\beq
\langle P_L \rangle=
-C \, \left ( 1+ {4\, \alpha_s\over 3 \, \pi}\right ) {\rm Re} \left (
\sqrt{E+i\Gamma \over m_t} \, \right ) ~,
\label{plc}
\eeq
which is the final result of this paper for the longitudinal
polarization of the top quark near the threshold. To repeat, the
coupling constant $\alpha_s$ in this expression is normalized at a short
distance scale $O(m_t)$ and the subsequent corrections are of the order
of $\alpha_s^2$, $\alpha_s v$, and $v^2$.

As a final remark, not related to the top quark polarization, but rather
related to the total cross section near the threshold, we note, that due
to the Coulomb renormalization the ratio of the cross section for
production of the quark pair in the $P$ wave to that in the $S$ wave is
not vanishing at  the threshold, i.e. at $v \to 0$:
\beq
\lim_{v \to 0} \left | {v \, P \over S} \right |^2 = {4 \over 9} \,
\alpha_s^2~~,
\eeq
so that the suppression of the $P$ wave cross section relative to the
$S$ wave exactly at the threshold is determined by $\alpha_s^2$, rather
than by $v^2$ {\footnote{This threshold behavior of the $P$ wave cross
section, also previously pointed out for the heavy scalar ( stop)
production$^{\cite{ifk}}$, is, of course, a consequence of the general
property of the Coulomb renormalization of the wave function with an
orbital momentum $L$ near the origin (see e.g. in the textbook
\cite{ll}): the threshold factor $v^{2L+1}$ is replaced by
$\pi \, a^{2L+1}$ with $a$ in our case being $a=2 \alpha_s/3$.}.
Thus, formally, in a calculation of the total cross section near the
threshold in order $\alpha_s^2$ one should also add the contribution of
the axial current of the top quark. The net effect of this contribution
is, however, quite small: the axial current cross section relative to
the vector current one amounts to
\beq
{\sigma_A \over \sigma_V}=  {2 \, D_A \over 3 \, D_V} \,{4 \over 9}  \,
\alpha_s^2 \approx 0.20 \, {4 \over 9}  \, \alpha_s^2~~,
\eeq
Naturally, the relevant scale for the $\alpha_s$ in this behavior is the
Coulomb one: $2 \alpha_s m_t/3$.

We are thankful to A.Hoang, M. Je\.zabek and V. Khoze for enlightening
comments.

This work is supported in part by DOE under the grant number
DE-FG02-94ER40823.

\end{document}